\documentclass{aastex}
\usepackage{graphicx}
\usepackage{amsmath}

\def \aj              {{\it Astron. J.}}

\def \aap             {{\it Astron. Astrophys.}}

\def \apj             {{\it Ap. J.}}
\def \apjl            {{\it Ap. J. Lett.}}

\def \apjs            {{\it Ap. J., Supp. Ser.}}

\def \mnras           {{\it MNRAS}}
\def \nat             {{\it Nature}}
\def \pasp            {{\it Pub. A. S. P.}}

\DeclareMathOperator*{\Median}{median}

\begin{document}


\title{The eShel Spectrograph: A Radial-Velocity Tool at the Wise Observatory} 

\author{M. Engel,  S. Shahaf , T. Mazeh}
%
%
\affil{
School of Physics and Astronomy, Tel Aviv University, Tel Aviv, Israel
}
\received{December 13,2016}
\accepted{March 9,2017}

%

\begin{abstract}
The eShel, an off-the-shelf, fiber-fed echelle spectrograph ($R \approx 10,000$), was installed on the 1m telescope at the Wise observatory in Israel.
We report the installation of the multi-order spectrograph, and describe our pipeline to extract stellar radial velocity from the obtained spectra. We also introduce a new algorithm---UNICOR, to remove radial-velocity systematics that can appear in some of the observed orders.
We show that the system performance is close to the photon-noise limit for exposures with more than $10^7$ counts, with a precision that can get better than 200 m/s for F--K stars, for which the eShel spectral response is optimal.
This makes the eShel at Wise a useful tool for studying spectroscopic binaries brighter than   $m_V=11$.
We demonstrate this capability with orbital solutions of two binaries from  projects being performed at Wise.
\end{abstract}

\section{Introduction}

The  radial-velocity (RV) technique led to the discovery of many planets and brown dwarfs \citep[e.g.,][]{Latham89, MayorQueloz95,  Mayor2014}.
 The discoveries were made by high-resolution spectrographs that have reached recently a remarkable precision that can get below $1$ m/s  \cite[see review by][]{Fischer2016}.
However, RV measurements with a precision on the order of a few hundreds m/s can still be important for the study of spectroscopic binaries with stellar or brown dwarf companions, and to rule out transiting planet candidates by detecting their binary nature \citep[e.g.,][]{Kirk2016,Kiefer2016,Hallakoun2016,Ma2016}.

This paper describes the installation and performance of a new, off-the-shelf, medium-resolution ($R \approx 10,000$) echelle spectrograph---eShel, on the 1m telescope at the Wise Observatory, Israel.
Similar spectrographs were installed recently in two other observatories \citep{csak14,pribulla15}.
Such systems offer a fast and simple way to enhance the measurement capability of small to medium scale telescopes to a 100 m/s  RV measurement tool.

Section~2 gives a short description of the spectrograph, Section~3 presents its installation and characterization at the Wise observatory. Section~4 describes our UNICOR---RV extraction algorithm developed for the reduction of the eShel observations and Section~5 presents the eShel+UNICOR performance.
Finally, Section~6 brings two binary orbits obtained with the new eShel, demonstrating the capability of the system. Section~7 briefly summaries  the presentation of the system. 

\section{The eShel spectrograph system}

 The eShel is an off-the-shelf, fiber-fed echelle spectrograph manufactured by the French company Shelyak Instruments.\footnote{http://www.shelyak.com}
The system was acquired on 2013 and installed on the wide-field 1m Boller and Chivens telescope, an  F/7 Ritchey-Chretien reflector on an off-axis equatorial mount at the Wise Observatory\footnote{https://physics.tau.ac.il/astrophysics/wise\_observatory} (34$^{\circ}$45'48''E, 30$^{\circ}$35'45'' N, 875 m) in Israel. 
The Wise observatory site provides $\sim$270 available nights per year with average seeing of 2.5 arcsec.
  
\begin{figure}[h!]
\centering
 \includegraphics[scale=0.5]{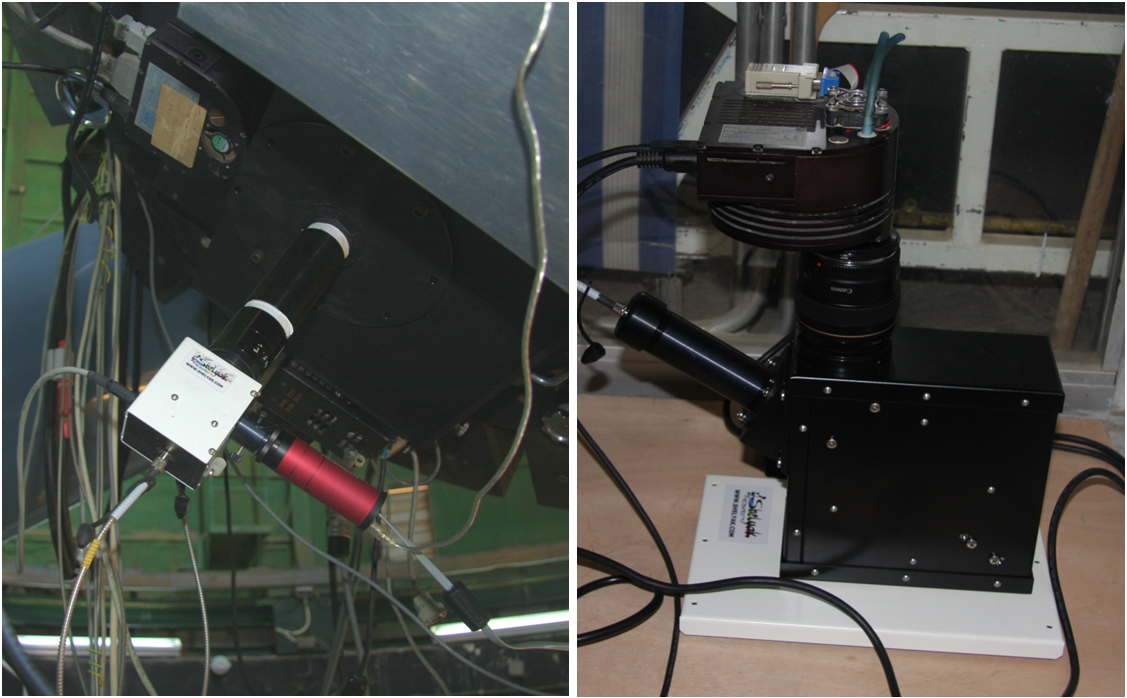}
 \caption{The eShel spectrograph (right) and the Fiber Injection and Guider Unit (FIGU) coupled to the telescope with a 0.63$\times$ focal reducer (left).}
 \label{eShel}
 \end{figure}

The spectrograph is fed by a $50 \mu m$ fiber (Fig.~\ref{eShel}).
The light from the fiber is collimated by F/5 optics on a 50$\times$25mm R2 blazed grating (blaze angle of $63.4^{\circ}$) with 79 lines/mm, illuminated in quasi Littrow configuration.
The light dispersed by the grating is reflected towards a prism cross disperser. An F/1.8 85mm Canon photographic lens focuses the echelle spectrum on an SBIG ST10XME CCD
 (2184$\times$1472  $6.8 \mu m$ pixels).
Orders 30--50 are used spanning a wavelength range of 4500--7600$\AA$.
The eShel is controlled and operated by the Audela\footnote{http://www.audela.org} open-source astronomy software, which also obtains the spectra from the CCD images.

\section{Installation at Wise}
\subsection{Coupling the eShel to the Telescope}

In fiber-fed spectrographs, the photons collected by the telescope aperture are focused on the fiber tip that acts as a limiting aperture (``slit'') of the spectrometer.
The size of the stellar disk image at the focal plane should be comparable to the size of the limiting aperture---the fiber-tip diameter.
The coupling efficiency between the telescope and the spectrograph is defined as the fraction of the photons reaching the focal plane that enter the fiber.
To achieve optimal coupling efficiency with the telescope a 0.63$\times$ focal reducer was introduced between the telescope and the FIGU (Fiber Injection and Guider Unit) to match the F/7 focal ratio of the telescope to the F/5 spectrograph.
With this focal reducer, the $50\mu m$ entrance aperture of the eShel is equivalent to 2.3 arcsec.
With the average 2.5 arcsec seeing at the Wise observatory site, about half of the photons reaching the focal plane are transmitted into the fiber.
Further reduction of the focal length to reduce the stellar image would result in losses at the F/5 optics of the spectrograph that will cancel the gain in the input end of the fiber.

The overall throughput of the eShel setup was measured from the ratio of the guiding CCD counts at the entrance of the spectrograph fiber to the total counts obtained in the spectra, summed over all the orders.
This yielded an average value of $3.4 \% \pm 1.6 \% $ over a sample of exposures of F,G,K stars taken in a time span of about 3 months with different seeing conditions.

\subsection{The Reduction Pipeline}
The eShel reduction pipeline is based on the eShel module Ver.~2.4 in Audela, written
%
%
by Michel Pujol and Christian Buil based on the algorithm outlined by \cite{Horne1986}.

\begin{figure}[h!]
\centering
 \includegraphics[scale=0.5]{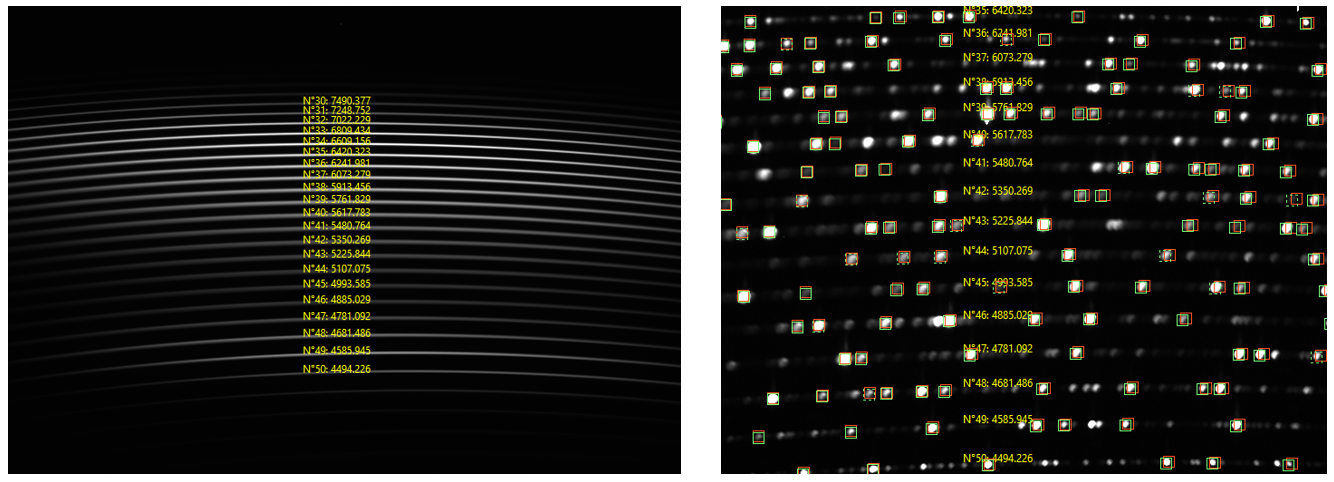}
 \caption{Left: Flat image from the eShel pipeline. Order numbers with the center wavelength of each order were identified by the software. The bottom orders are enhanced with blue LED illumination. Right: A magnified part of the processed ThAr calibration. Identified emission lines are marked: green squares---the nominal position of the line and red squares---the corrected position with the calibration correction applied.}
 \label{FlatThAr}
 \end{figure}
  \begin{figure}[h!]
\centering
 \includegraphics[scale=0.6]{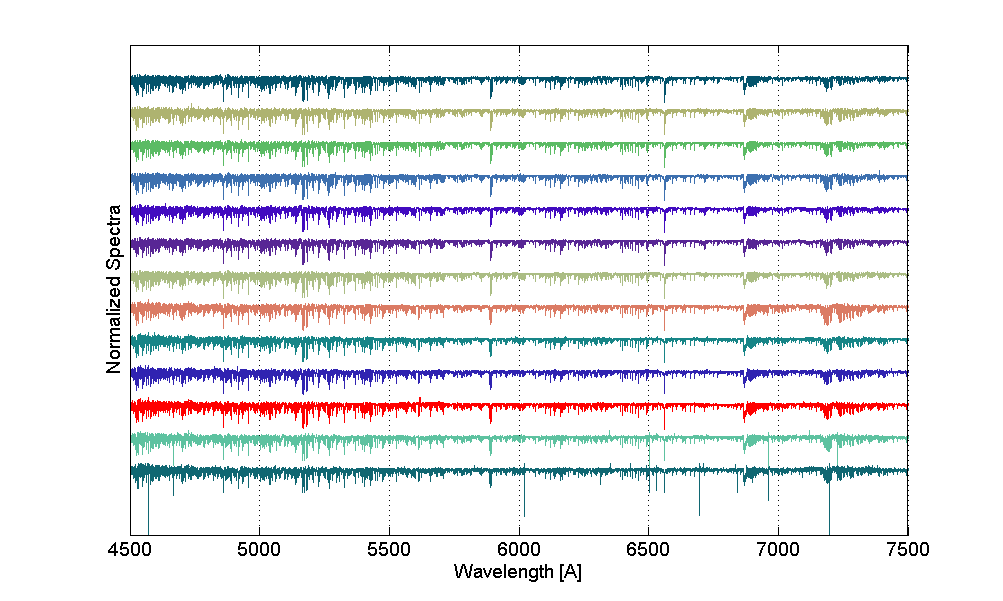}
 \caption{eShel spectra extraction pipeline result: a series of normalized
flattened spectra of HD 67767 taken during 2015–-2016. Comparing the
obtained spectrum to a G7V template, it is clear that the scatter in the blue part
of the figure is due to the lower signal coming from the star (not reflected here,
as the spectrum is flattened), and to the lower sensitivity of the system.}
 \label{HD67767Multi}
 \end{figure}

The reduction follows standard procedures of fiber-fed echelle spectrograph data processing \citep[e.g.,][]{Brahm2016}.
The pipeline uses reference images of dark, bias, continuum exposures of a tungsten lamp and flat images of tungsten lamp with added blue LED illumination to allow better tracing of the blue orders.
The tracing of the orders is done on bias-dark processed flat images.
The flat processing routine identifies the orders, fits a 5th-order polynomial to trace the arc image of the order on the CCD image and extracts the blaze profile for each order (Fig.~\ref{FlatThAr} left panel).

The wavelength calibration is based on exposures of low-pressure ThAr lamp, whose light is injected into the spectrograph input fiber and goes through the same optical path as the stellar light.
Before each science exposure, two 10-sec exposures of the ThAr lamp are taken and their sum is used for wavelength calibration. 
The calibration algorithm identifies the location of more than 300 ThAr lines over all orders and fits, for each order, a 3rd-order polynomial correction to the nominal spectrograph dispersion function (Fig.~\ref{FlatThAr} right panel).

The spectrum is extracted from the science exposures following the traces determined from the flat images.
Wavelength calibration is applied to the data and the raw order spectra are re-sampled on a uniform wavelength grid.
The final stage of the pipeline prepares the spectra for the RV extraction
by dividing  the spectra by the continuum to produce flat normalized spectrum for each order (Fig.~\ref{HD67767Multi}).

\subsection{eShel Spectral Response}

\begin{figure}[h!]
\centering
 \includegraphics[scale=0.6]{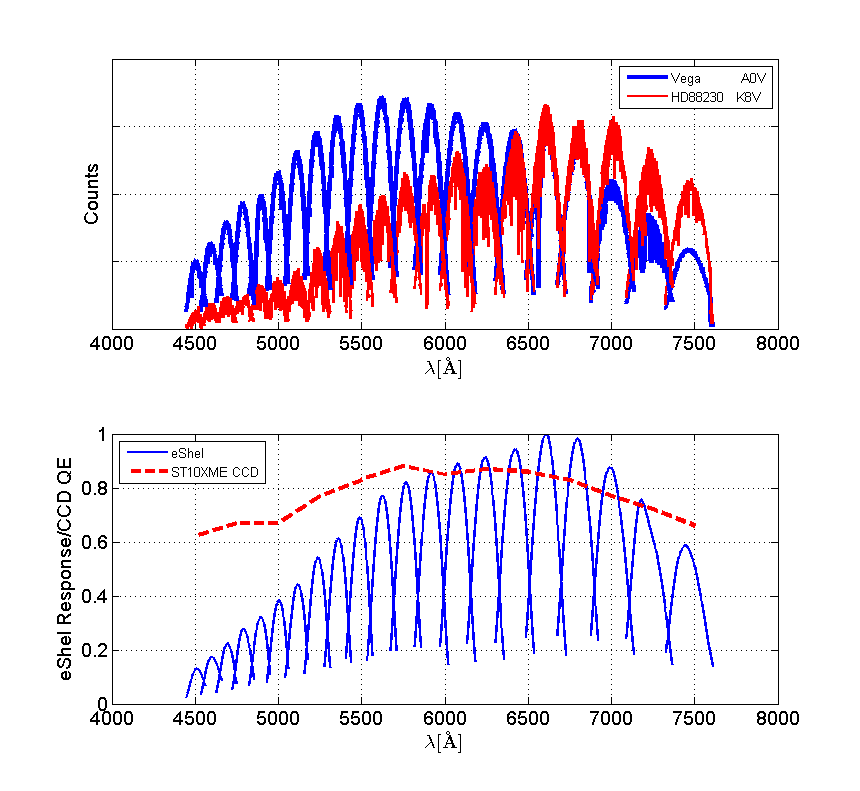}
 \caption{Top: CCD counts as function of wavelength in orders 30--50 of the eShel acquired for two stars of different spectral type: Vega A0V (blue) and HD 88230 K8V (red). The strong effect of the blaze function in each order and the global behavior of the blaze maxima reflect the effect of the global eShel system spectral response on the counts collected in each spectral bin. Bottom: eShel system (telescope-fiber-spectrograph-CCD) spectral response (blue). The throughput curve is normalized to the maximum value at $6600\AA$. The SBIG ST10XME CCD quantum efficiency (red) is given for reference.}
 \label{vegaHD88230sig_resp}
 \end{figure}

The spectrograph spectral response is defined as the ratio of $n(\lambda)$ (counts/s) collected at the CCD to the flux $\phi(\lambda)$ (photons/s)  incident on the telescope aperture in a certain wavelength $\lambda$.
Fig.~\ref{vegaHD88230sig_resp} (top) shows the CCD counts collected for two different spectral-type stars, Vega (A0V) and HD 88230 (K8V).
The strong variation in each order (blaze function) and the global behavior of the order maxima can be clearly seen in the plot.
As described in the previous section, the blaze function for each order is extracted from flat reference images.
The global spectral dependence of the response is a product of the atmospheric transmission, telescope optics and fiber transmission, the blazed grating diffraction efficiency and the CCD quantum efficiency.
  
During the commissioning phase of the spectrograph, several consecutive  exposures of Vega were taken.
The raw signal was then divided by the spectral flux of an A0V star   \citep{Pickles1998} and by the calculated atmospheric transmission profile.
Fig.~\ref{vegaHD88230sig_resp} (bottom) shows the derived eShel spectral response with the SBIG ST10XME CCD quantum efficiency for reference.
The spectral response is skewed towards the red part of the visible spectrum, making  eShel performance best for F--K stars.

\newpage

\subsection{Wavelength Calibration Stability}
As described above, wavelength calibration is based on ThAr exposures, where the positions of known spectral lines are used to generate wavelength calibration.
Position changes of the ThAr lines during the night due to instrumental variations cause  systematic errors which show as RV drifts. 
The main cause for such variations are temperature gradients in the immediate environment of the spectrogtraph.
To minimize these effects we have put the spectrograph in a thermally insulated box.
In addition, we keep track of the calibration stability as part of the pipeline by tracking positions on the CCD of a sample of 17 ThAr spectral lines in different orders.

\begin{figure}[h!]
\centering
 \includegraphics[scale=0.6]{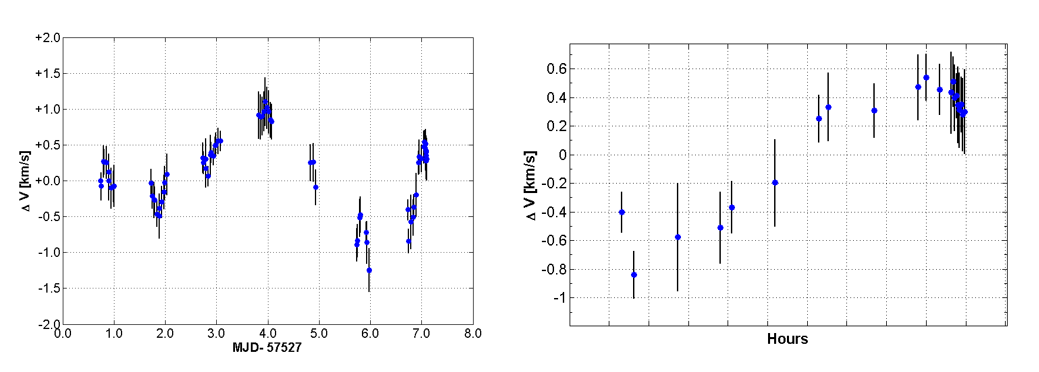}
 \caption{Drifts of ThAr calibration line positions on the CCD during 7 consecutive nights (left) and a zoom on the last night (right). Each point is the median of the drifts of 17 lines in a specific calibration relative to the first calibration. The size of the error bar is the scatter in the drifts of the different lines in the same calibration.}
 \label{calib_stability}
 \end{figure}
 
Fig.~\ref{calib_stability} shows the median drift of the position on the CCD of the 17 spectral lines relative to their initial position for 7 consecutive nights (left) and a zoom on the last night (right),
expressed in terms of $\Delta V=c \frac{\Delta \lambda}{\lambda}$ km/s.
The error bars represent the scatter of the drift of the different calibration lines.
 The scatter of line positions in consecutive calibrations is of the order of $100$ m/s, equivalent to a scatter of $\sim 0.02$ pixel on the CCD.
 This scatter sets a lower limit to the spectrograph precision.
 
The span of the nightly drift, which can be as high as 1--2 km/s, may add a drift of the order of 300 m/s for 1 hr exposures. This drift, however, can be compensated by monitoring RV standard stars and telluric lines.

\subsection{eShel spectra}

To conclude this section we show in Figure \ref{HIP116085_HD58728_spectra} one order of the eShel spectra of two stars, HD 221354---a standard star (see below) and HD 58728---a double-lined spectroscopic binary \citep{Abt_Levy1976} that we are following (see Section 6).

\begin{figure}[h!]
\centering
 \includegraphics[scale=0.5]{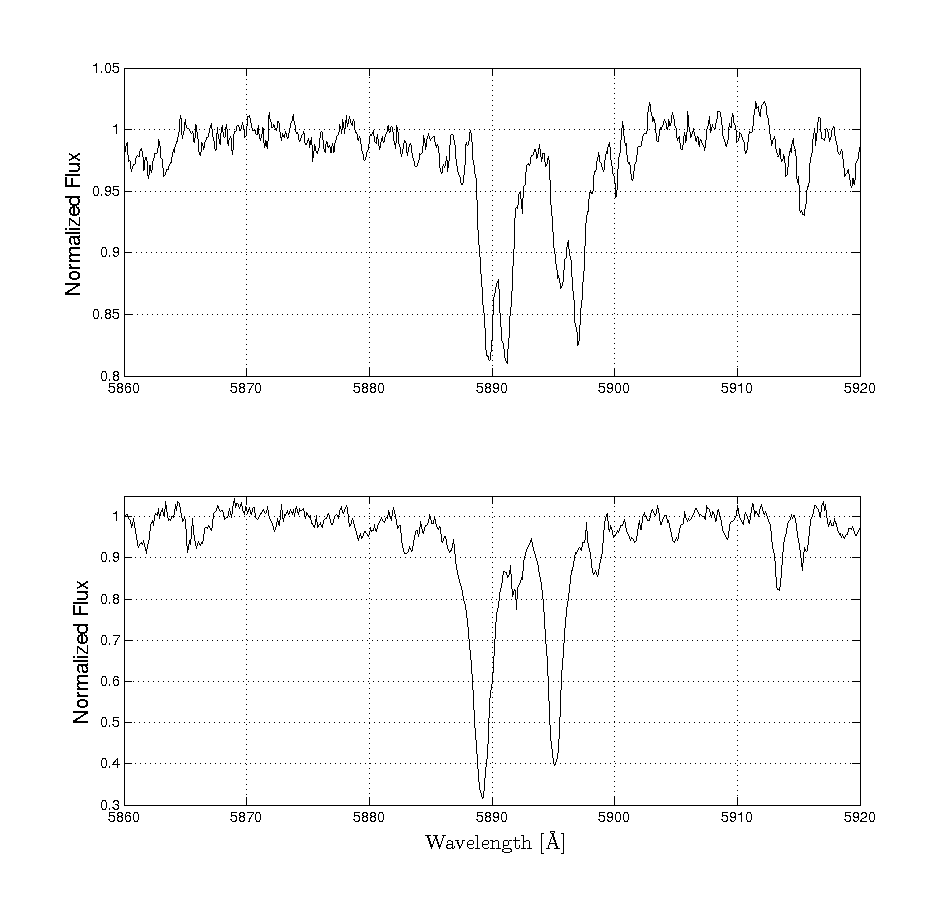}
 \caption{One order of the eShel spectra with the sodium D-line doublet. Lower panel shows the normalized spectra of the RV standrad HD 221354 (see Fig.~4). Upper panel shows the spectra of HD 58728, a double lined spectroscopic binary. 
}
 \label{HIP116085_HD58728_spectra}
 \end{figure}
%
%
%

\section{UNICOR: RV extraction algorithm}
%

The UNICOR algorithm  extracts RVs from single-lined multi-order stellar spectra by cross correlating  the observed stellar spectra against a given template \citep{tonry79}. In most cases
UNICOR templates are based on synthetic models taken from the PHOENIX\footnote{G\"{o}ttingen Spectral Library, \url{http://phoenix.astro.physik.uni-goettingen.de}} library \citep{husser2013}.
The PHOENIX templates are adapted to the eShel resolution by convolution with a Gaussian profile,
where the mean width at half maximum was experimentally found to be $0.9 \mbox{\normalfont\AA}$. 
Additionally, a rotational broadening profile \citep{gray1992} is applied to account for the stellar rotation $v\,sin(i)$.

%

In cases where the spectral properties of the star are unknown, a grid search of the library is preformed over $T_{eff}$, $log(g)$, [Fe/H] and $v\,sin i$ parameter space.
The maximum value of the cross-correlation function (CCF) serves as a goodness-of-fit measure, 
thus the template with highest median CCF peak (over all orders and observations) is chosen.

\subsection{RV derivation algorithm outline}

The UNICOR algorithm derives RVs for  each exposure by locating the peak of the cross correlation of each order as a function of the template shift, and then averaging over the shifts of the different orders to get the RV of that exposure. The heart of the algorithm is to identify and then remove systematic effects found in some of the orders.

Let us denote by $V_{i,j}$ the RV obtained per order $i$ for each exposure $j$. 
We assume that
\begin{equation}  
V_{i,j}=V_j+\epsilon_{i,j} \ ,
\end{equation}
where $\epsilon_{i,j}$ is the error of the $i$-th order at the  $j$-th exposure, and $V_j$ is the true stellar velocity at the time of the $j$-th exposure.  
The error matrix includes noise, and may also contain systematic errors of unknown nature.
Some orders might be contaminated by telluric lines or suffer from wavelength calibration errors.
 The UNICOR algorithm we developed aims to improve the derivation of  the RVs and their errors  by identifying and correcting the systematic errors at different orders.
 As we do not know the true values of the stellar RVs, $V_j$, the algorithm is iterative.
 
Denote $V_{i,j}^0\equiv V_{i,j}$ in the zero (initial) iteration.
The derivation starts with a subset of orders, $\{I_0\}$, selected for the analysis by eliminating orders governed by telluric lines and orders with low signal-to-noise ratio.
 An initial estimate of the stellar velocity and its uncertainty is obtained for each exposure by taking the median and median absolute deviation of this subset respectively.
Hence
\begin{gather} \label{eq:<RV_0>}
V_{j}^0= \Median_{i \in I_0} \{V_{i,j}^0\} \ ,
\end{gather}
where the medians are taken over all orders in $I_0$. 
The zero-iteration error matrix is
\begin{equation}
\epsilon_{i,j}^0 = V_{i,j}^0 - V_j^0 \ .
\end{equation}
With no systematic errors, we expect each row $i$ in the $\epsilon_{i,j}$ matrix, which corresponds to the  $i$-th order, to be scattered evenly around zero.
An initial estimate of the systematic shift of an order $i$ is therefore given by
\begin{equation}
\delta_i^0=\Median_{All\,\,j}\{\epsilon_{i,j}^0\} \ . 
\end{equation}
The scatter of order $i$, at the current iteration, is derived from the median absolute deviation of the corrected error matrix,
\begin{equation}
\sigma_i^0 = 1.48 \cdot \Median_{All\,\,j}\{| \epsilon_{i,j}^0 - \delta_i^0|\}  \ .
\end{equation}
Significant systematic shifts, denoted $\widetilde{\delta_i}^0$, 
are defined by the criterion $|\delta_i^0| > C\cdot \sigma_i^0/\sqrt{n_{obs}}$, 
where $n_{obs}$ is the number of observations and $C$ is a shift significance parameter.
If the criterion is met, we consider the shift to be significant and
$\widetilde{\delta_i}^0 =\delta_i^0 $, otherwise $\widetilde{\delta_i}^0=0$.
Significant shifts are removed from the velocity matrix $V_{i,j}$. 
From the corrected velocity matrix, 
\begin{equation}
	V^1_{i,j} \equiv V_{i,j}^0 - \widetilde{\delta_i}^0 \ , 
\end{equation}
a new set of stellar velocities is derived.
A new set of orders $\{I_1\} \subseteq \{I_0\}$ 
is generated by rejecting orders with $\sigma^1_i$ larger than a predetermined threshold, $\sigma_{th}$.
\begin{gather} \label{eq:<RV_1>}
V_{j}^1= \Median_{i \in I_1} \{V_{i,j}^1\} \ ,  \\ \nonumber
\sigma_j^1 = 1.48 \cdot \Median_{i\in I_1}\{|V_{i,j}^1 - V_{j}^1|\} \ .
\end{gather}
Generally, for the $k+1$ iteration,  
\begin{gather} 	\label{eq:<RV> k iter}
V^{k+1}_{i,j} \equiv V_{i,j}^{k} - \widetilde{\delta_i}^{k} \ , \\ \nonumber
\epsilon_{i,j}^{k+1} = V_{i,j}^{k+1} - V_j^{k+1} \ .  
\end{gather}
Before repeating the velocity derivation process, a new subset of orders, $\{I_{k+1}\}$, is selected from the initial set $\{I_0\}$ 
 by rejecting orders with $\sigma^k_i$ larger than the predetermined threshold $\sigma_{th}$.
This process is repeated until the stellar RV values $V_j$ converge. 
The products of the algorithm are the velocities and errors of the final iteration:
\begin{gather} \label{eq:<RV_final>}
V_j = V_{j}^{final} \ ,  \\ \nonumber
\sigma_j = \sigma_{j}^{final} \ .
\end{gather}
\subsection{UNICOR performance demonstration}
As a test for UNICOR, 41 exposures of the RV standard HD 221354 (V = 6.74 K0V) were taken over a time span of two months. 
The initial set of orders, $\{I_0\}$, consisted of 16 orders that span the spectral range from  $4534$ to  $6726 \textrm{\AA}$, after removal of orders dominated by telluric lines.
Ten iterations of the algorithm were preformed with $\sigma_{th}=1$ km/s and $C=1$.
Convergence was tested by calculating the median change in velocity (over all exposures) between consecutive iterations.

\begin{figure}[h!]
\centering
 \includegraphics[scale=0.5]{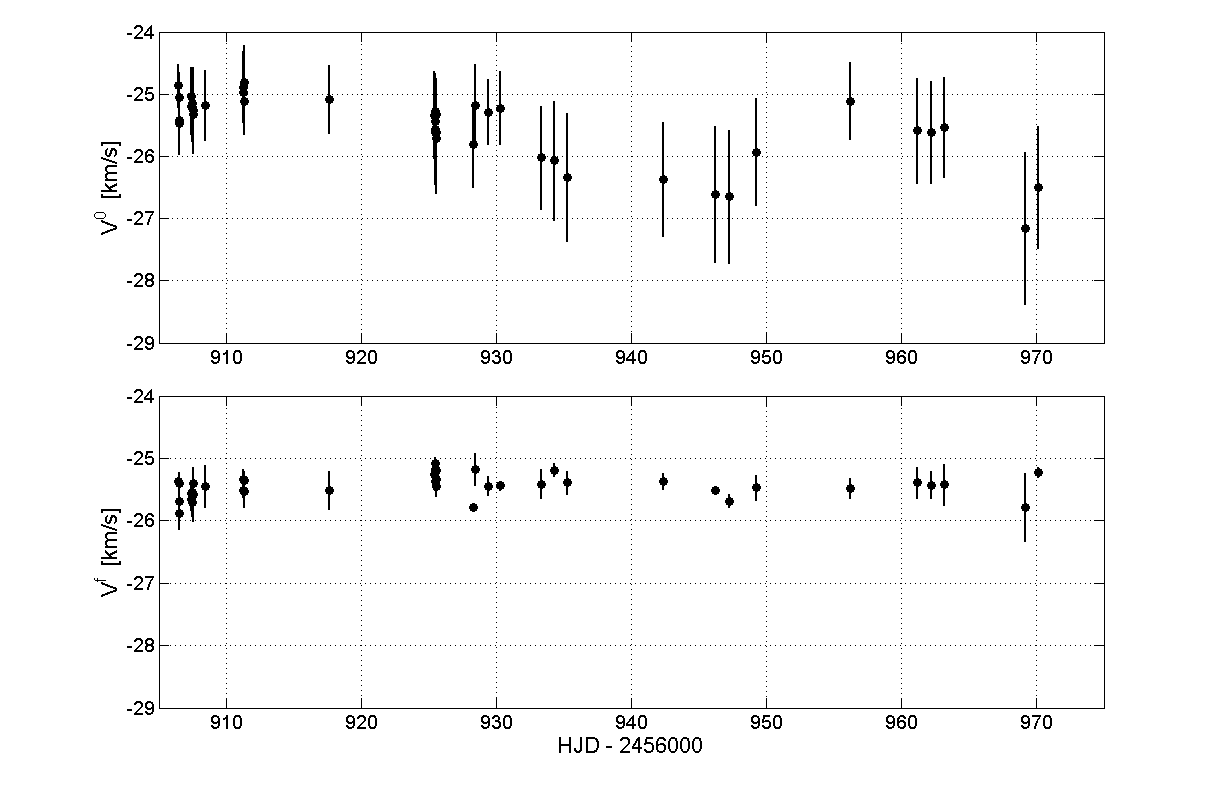}
 \caption{RV measurements of HD 221354, before (upper panel) and after (lower panel) systematics removal by UNICOR.
}
 \label{unicor: HIP116085}
 \end{figure}

As shown in Figure~\ref{unicor: HIP116085} (upper panel), the initial velocity estimates, $V^0_j$, were contaminated by systematic errors with a scatter of $380$ m/s.
The velocities obtained in the final iteration yielded a scatter of $130$ m/s around the median velocity and had much less systematic velocity drifts (Figure~\ref{unicor: HIP116085} lower panel).
The median of $\sigma_j$ in the final UNICOR iteration was 162 m/s, similar to the observed scatter. 

The stellar velocity derived,  $-25.43 \pm 0.13$ km/s, was slightly lower, but within  $2\sigma$, of the value of $-25.113\pm 0.122$ km/s, given in the RV standard list of \citet{Chubak2012}.
The difference could be due either to
some statistical fluctuation or a zero-point difference between
the two instruments.
\section{eShel Performance}

\subsection{Precision vs. Total Collected Counts}

The performance of the spectrograph with its pipeline as an RV tool is characterized by the number of photons needed to achieve a given RV precision.
\cite{Bouchy01} showed that the photon-noise-limited RV precision achievable by a spectrograph  follows the  expression:
\begin{equation} \label{eq:bouchy1}
\sigma=\frac{c}{Q \cdot \sqrt{N_{e^-}}} \; .
\end{equation}
In this expression $\sigma$ is the uncertainty of the measured RV \cite[$\delta V_{RMS}$ in ][]{Bouchy01}, c is the speed of light, $N_{e^-}$ is the number of counts (photoelectrons) that generate the spectrum over all the spectral range, for a specific exposure, and Q is a factor that describes the amount of spectral information in the spectrum.
The Q values range from few 100's for $R\sim 10,000$ spectrographs to 10,000's for high resolution spectrographs, depending on the spectrograph resolution, spectral range,  and the richness of lines in the stellar spectrum, i.e.~the spectral type of the star.
The performance of the eShel and its pipeline can be mapped by plotting the typical error as a function of the total counts.

In order to assess the performance of the eShel at Wise, we measured during various runs three RV standards: HD 67767 (V=5.7 G7V), HD 221354 (V=6.7 K0V) and HD 88230 (V=6.6 K8V), taken from \cite{Chubak2012}, with different exposure times. 
The measured data set spanned three years, starting at the beginning of the eShel operation.
This gave us spectra with varying levels of SNR and therefore varying level of internal error estimate.

%
\begin{figure}[h!]
 \includegraphics[scale=0.8]{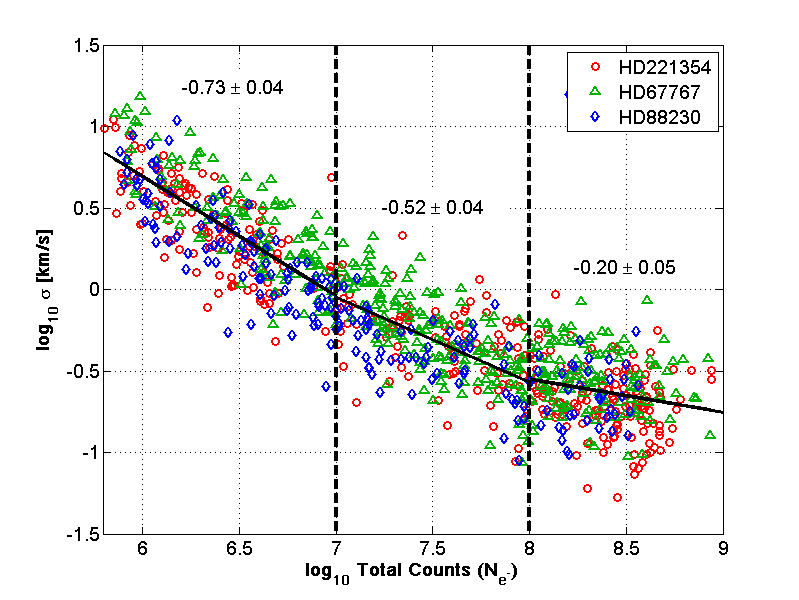}
 \caption{RV uncertainty vs.~total counts for HD 221354 (circles), HD 67767 (triangles), HD 88230 (diamonds) plotted on log-log scale. The figure is divided into three ranges, with different linear slopes shown as black lines.  
At the range of $10^7$--$10^8$ counts the errors are dominated by the photon-noise, with a slope of $-0.52 \pm 0.04$, a value close to the one expected from Eq.~\ref{eq:bouchy1}.
Below $10^7$ counts, the slope is higher, at a value of $-0.73 \pm 0.04$, because the CCD noise contribution becomes significant.
Above $10^8$ counts the slope is quite small, probably dominated by higher contribution of the systematic effects.
}
 \label{dV_vs_total_counts}
 \end{figure}

Figure~\ref{dV_vs_total_counts}, plotted on a log scale, shows the derived errors as a function of the total counts for the observations of the three RV standards.
For each exposure, V and $\sigma$ were derived by UNICOR and the total counts were summed over all orders.
The figure shows three regions. 
At the range of $10^7$--$10^8$ counts the errors are dominated by the photon-noise, with a slope of $-0.52 \pm 0.04$, a value close to the one expected from Eq.~\ref{eq:bouchy1}.
Below $10^7$ counts, the slope is higher, at a value of $-0.73 \pm 0.04$,  because at this region the CCD (readout+dark) noise contribution becomes significant. At this level of counts the eShel precision is below the photon noise limit and deteriorates faster with lower counts. 
Above $10^8$ counts the slope is quite small, probably dominated by higher contribution of the systematic effects.

The median value of counts for the eShel at Wise, normalized to $m_V=0$, air mass of 1 and seeing of 2.5", was found to be $(1 \pm 0.5)\cdot 10^8$ counts per second. 
This means that to be well in the photon noise region, the required exposure time is given by:
  \begin{equation} 
t_{exp} \geq 0.3 \cdot 2.5^{m_V} \; sec,
\end{equation}
for a star with visual magnitude $m_V$. 

\subsection{External Error vs. Internal Error}

To compare our error estimate, $\sigma$, with some estimation of the external error, 
$\sigma_{ext}$, we used the same data set of the three RV standards as above, and obtained for each velocity the deviation from the corresponding median velocity
\begin{equation} \label{eq:DeltaV}
\Delta V_j=|V_j - \Median_{ All\, j} (V_j)| \; .
\end{equation}
Binning the data sets of the three RV standards into logarithmically sized bins of the internal error $\sigma$, we obtained $\sigma_{ext}$ as the scatter of the velocity deviations, estimated by the 68-th percentile of the $\Delta V_j$ distribution in each bin. 

 \begin{figure}[h!]
\centering
 \includegraphics[scale=0.8]{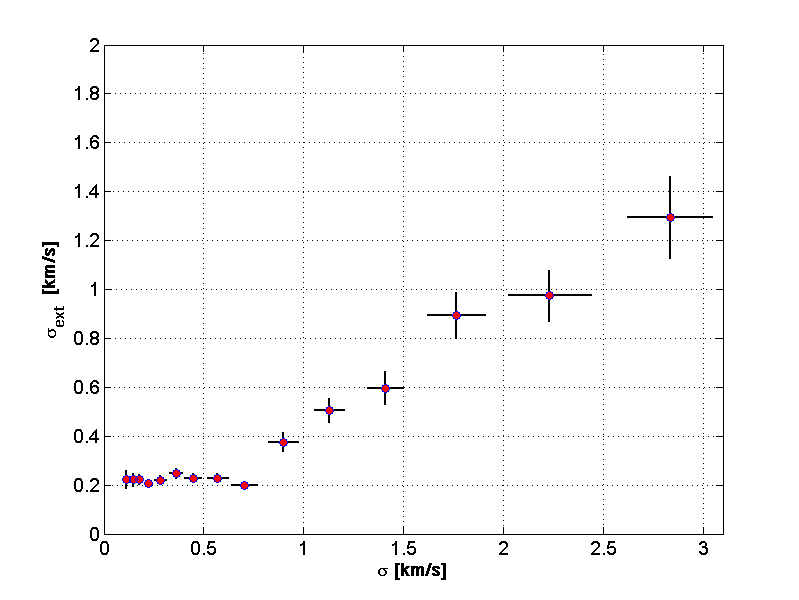}
 \caption{External error as a function of the internal error derived by UNICOR. The data of the three RV standard stars are binned in equal bins of the internal error $\sigma$. 
The typical external error is derived from the scatter of the velocity deviations (see text). 
The horizontal error bars show the scatter of $\sigma$ values in the bin and the vertical error bars show the {\it expected} error  of the external error estimate.}
 \label{deltaV_vs_sigmaV}
 \end{figure} 

Figure~\ref{deltaV_vs_sigmaV} shows a plot of $\sigma_{ext}$ as a function of $\sigma$. 
The horizontal error bars show the scatter of the $\sigma$ values in the bin, and the vertical error bars show the expected error in the derived scatter in that bin, based on the square root of the number of points in the bin. 
Figure~\ref{deltaV_vs_sigmaV} shows a linear correlation between the two at a range of 0.7--3 km/s. For the sample of the three bright RV standards, the internal error in this range is about twice the external error.
The left-hand side of the plot shows that the external errors of these measurements reached a limit around 200 m/s for internal errors below 700 m/s.

\newpage

\section{Two eShel Projects at Wise}

In this section we present two examples of orbital solutions obtained for spectroscopic binaries in two projects being performed at Wise.
A detailed report on the projects will be published elsewhere.
\subsection{New Orbits for Known Spectroscopic Binaries}
We are performing a small-scale survey of spectroscopic binaries known already for a few decades, to find out if their orbital elements changed during these years \citep{MM87}.
Such changes, if exist, could be caused by dynamical interaction with a third unseen distant companion orbiting the binary system \citep{MazehShaham1976,MazehShaham1979,Kiseleva1998,FabryckyTremaine2007}.

\begin{figure}[h!]
 \centering
 \includegraphics[scale=0.8]{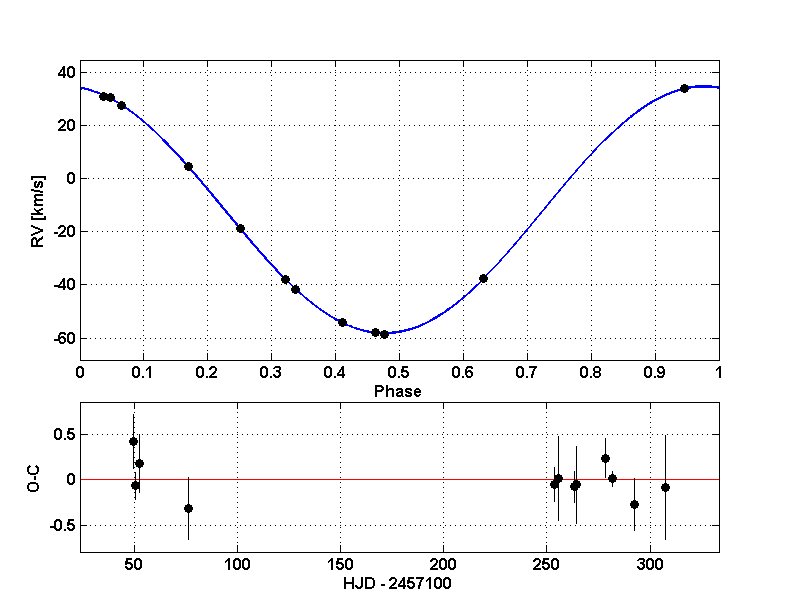}
 \caption{HD 95363 Solution based on the eShel data (top panel) and the residuals (bottom panel). The error bars in the phase plot are too small to be seen.}
 \label{HD95363_sol}
 \end{figure}

Figure~\ref{HD95363_sol} shows an orbital solution for HD 95363 (V=7.95 F7V) based on 12 eShel exposures taken during 2015.
The same system was measured by \cite{imbert72}, who obtained 14 RVs of this system with the coude spectrographs of the 1.93m and 1.52m telescopes at the Observatoire de Haute Provence (OHP).
A new orbital solution was calculated based on 13 RVs (one measurement was discarded as outlier).
We took the errors of Imbert RVs to be proportional to $1/w$, where $w$ is the weight given for each RV measurement.
Table~\ref{tab:Table1} compares the orbital elements derived from our eShel velocities with the old data set, assuming a circular orbit. 
The external error $\sigma_{ext}$ of the velocities around the fitted model is 
 220 m/s while the median value for $\sigma$ from UNICOR is 250 m/s, in agreement to Figure~\ref{deltaV_vs_sigmaV}.
 
\begin{table}[h!]
\centering
\begin{tabular}{lll}

 HD 95363:& eShel Wise (2015) & OHP (1971)  \\
\hline
Period  & $7.84912 \pm 0.00018$ d & $7.8491 \pm 0.0011$ d\\
T & $2457297.0688 \pm 0.0007$ HJD &  $2440820.05 \pm 0.02$ HJD\\
K  & $46.530\pm 0.094$ km/s & $46.19 \pm 0.58$ km/s  \\
$V_0$ & $-11.684 \pm 0.055$ km/s & $-13.53 \pm 0.39$ km/s\\
e &  $0 \quad (fixed)$ &  $0 \quad (fixed)$  \\

 $\sigma_{ext}$ & $220$ m/s & $1180$ m/s \\
N           & 12 & 13 \\

\end{tabular}
\caption{Orbital elements of HD 95363 derived from  eShel measurements at Wise (2015)  compared to elements derived from RV measurements by Imbert at OHP (1971).}
\label{tab:Table1}
\end{table}

The period and K, the semi-amplitude, of the two solutions agree within their errors.
The systemic velocities $V_0$ differ, probably because of a shift of the RV zero point of the two observatories.

The eShel observations do not show any change in the RV amplitude of the orbital motion, as the difference $\Delta K= 0.34 \pm 0.59$ is not significant.
The full results of this project will be reported in detail in a forthcoming paper (Engel et al. in preparation).

\subsection{Follow-up measurements of non-eclipsing binaries discovered by the BEER algorithm}
The recent $CoRoT$ \citep{Auvergne2009} and $Kepler$ \citep{Borucki2003} space missions provided stellar light curves with precision in the range of several tens of ppms.
\cite{LoebGaudi2003} and \cite{Zucker2007} showed that this precision allows identifying non-eclipsing planets and binaries by detecting  the BEaming, Ellipsoidal and Reflection (BEER) effects.
The beaming effect is a relativistic Doppler effect that causes a change in the stellar flux that is proportional to its radial velocity.
The ellipsoidal effect modulates the flux of the system due to the ellipsoidal shape of the primary and secondary stars whose major axes are along the line connecting them, rotating with the binary period.
The flux is proportional to the projected star area on the line of sight, causing modulation with half the binary period. The reflection effect modulates the flux due to the reflected radiation from the secondary when it is on the farther side (opposing) of the primary.

\begin{figure}[h!]
 \centering
 \includegraphics[scale=0.8]{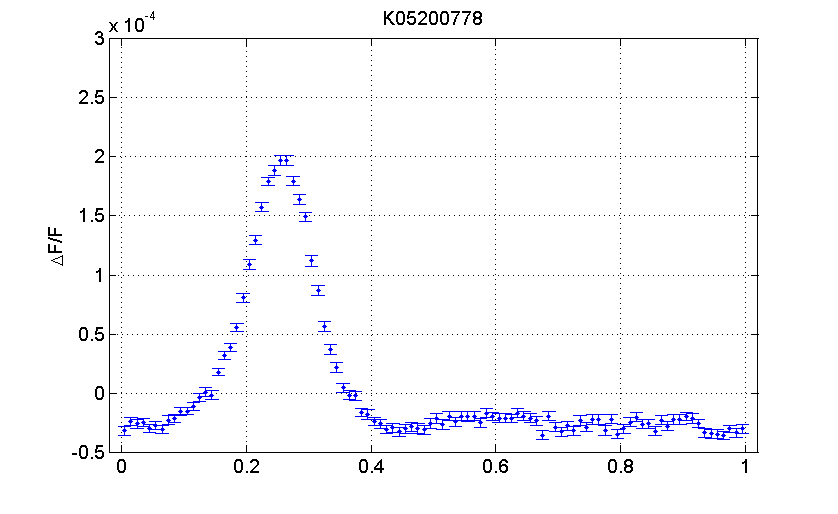}
 \caption{Light curve of K05200778, folded on a period of 16.35d found by the BEER analysis.
Each point represents binned data, where the bin size is 1/100 of the period.
The error bars present the scatter in the bin.
Note that the amplitude of the flux variation is of the order of $10^{-4}$.}
 \label{K05200778_lc}
 \end{figure}
%

The three effects modulate the light curve of a binary system differently, thus allowing to derive information on the orbital elements of the system.
\cite{Faigler2011} developed an algorithm to search for non-eclipsing binaries in the $Kepler$ and $CoRoT$ light curves, leading to the detection of non-eclipsing binaries and planets \citep{Faigler2012,Faigler2013,Faigler2015b,Tal-Or2015}.


At Wise we are following a few bright Kepler eccentric candidates found by the BEER algorithm.
One example is the $Kepler$ star K05200778 (V=9.5, $T_{eff}=5800$K, log g=4.0 [Fe/H]=0.0).
The algorithm found a strong periodic modulation with a period of 16.35d.
Figure~\ref{K05200778_lc} shows a folded and binned light curve of K05200778, where the data was  folded with the period found and binned into bins of 1/100 of the period. 
The light curve demonstrates an "eccentricity pulse" which appears in binary systems with significant eccentricity \citep{Dong2013}.
The "eccentricity pulse" is caused by the ellipsoidal effect when the two binary components go through periastron for systems that the line connecting the stars at periastron is close to be perpendicular to the line of sight.
The predicted orbital configuration of this binary candidate from the BEER analysis is of an eccentric binary with a period of 16.35d and $\omega$ close to zero.

\begin{figure}[h!]
 \centering
 \includegraphics[scale=0.8]{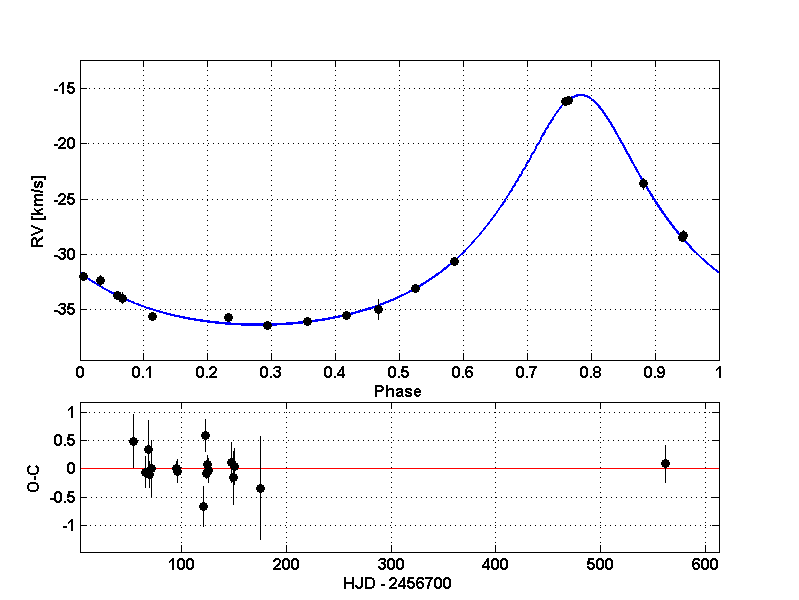}
 \caption{K05200778 Solution (top) and the residuals (bottom).}
  \label{K05200778_sol}
 \end{figure}

In the summer of 2013 and 2014, we performed follow-up observations of K05200778 with the eShel spectrograph.
The orbital elements (Table \ref{tab:Table2}) were derived from 16 exposures (Figure~\ref{K05200778_sol}).

 \begin{table}[h!]
\centering
\begin{tabular}{lll}
 K05200778& eShel Orbital Elements \\
\hline
Period  & $16.4032 \pm 0.0082$ d \\
T & $ 2457044.70 \pm 0.23  $ HJD\\
K  & $10.37\pm 0.21$ km/s \\
$V_0$ & $-30.109 \pm 0.086$ km/s  \\
e &  $0.398 \pm 0.013$ \\
$\omega$ & $0.8^o \pm 2.7^o$ \\
$\sigma_{ext}$  & $420$ m/s \\
N & 16   \\
\end{tabular}
\caption{Orbital elements of K05200778 derived from  eShel measurements}
\label{tab:Table2}
\end{table}

The derived period from the eShel measurements was $16.40$d, similar to the BEER period.
The solution is clearly eccentric with eccentricity of $0.4$ and $\omega$ was found indeed to be close to $0^0$, as predicted by the BEER analysis. Apparently, the photometric phase was shifted by half a period relative to the RV orbit. 
The internal error was 315 m/s  while the external error was 420 m/s, somewhat higher than expected  from Figure~5.
The full results of this project will be reported in detail elsewhere (Engel et al. in preparation).

\section{Summary}

This paper describes the installation of the eShel on the 1m telescope at the Wise observatory, including the optics, the CCD and the response of the system. We briefly overview the software of the pipeline, including UNICOR, a new algorithm to remove systematic effects in different orders. 

Analysis of measurements of RV standard stars shows that a photon noise limited precision of 200 m/s can be achieved for bright G and K stars.
With the 1m telescope at the Wise observatory, the eShel can be used to measure targets down to $m_V=11$.
The external error, as manifested by the scatter of the RVs around the fitted models, was shown to be correlated to the precision calculated by UNICOR.
The external error converges to a minimum of 200 m/s, a minimum that is limited by the stability of the eShel spectrograph.
These results are similar to previous reports from the other eShel installations \citep{csak14,pribulla15}.

We demonstrate the performance of the system in two projects: a survey of spectroscopic binaries to find variations in their orbital elements and follow-up of BEER non-eclipsing binary candidates, showing that the Wise eShel can be a valuable tool for the study of bright spectroscopic binaries.

\begin{acknowledgements}
We deeply thank Prof.~Dan Maoz, the Wise observatory director, for generous time allocation, and Shai Kaspi, Ezra Mashaal and Sami Ben-Gigi from the Wise Observatory team for their help and continued support during the eShel commissioning and afterwards.
Without their support this project could not have been realized.
We thank Fran\c{c}ois Cochard and his team from Shelyak Instruments for the continuous support with the eShel installation and maintenance.

 This research has received funding from the European Research Council under the EU’s Seventh Framework Programme (FP7/(2007-2013)/ERC Grant Agreement No.~291352), the ISRAEL SCIENCE FOUNDATION (grant No.~1423/11), and the Israeli Centers for Research Excellence (I-CORE, grant No.~1829/12).
\end{acknowledgements}

%
%

  \newpage

%


\begin{thebibliography}{}

\bibitem[Abt \& Levy(1976)]{Abt_Levy1976} Abt, H.~A., \& Levy, S.~G.\ 1976, \apjs, 30, 273 



\bibitem[Auvergne et al.(2009)]{Auvergne2009} Auvergne, M., Bodin, P., Boisnard, L., et al.\ 2009, \aap, 506, 411 

\bibitem[Brahm et al.(2016)]{Brahm2016} Brahm, R., Jord{\'a}n, A., \& Espinoza, N.\ 2016, arXiv:1609.02279 



\bibitem[Borucki et al.(2003)]{Borucki2003} Borucki, W.~J., Koch, D., Basri, G., et al.\ 2003, Earths: DARWIN/TPF and the Search for Extrasolar Terrestrial Planets, 539, 69 


\bibitem[Bouchy et al.(2001)]{Bouchy01} Bouchy, F., Pepe, F., \& Queloz, D.\ 2001, \aap, 374, 733 


\bibitem[Chubak et al.(2012)]{Chubak2012} Chubak, C., Marcy, G., Fischer, D.~A., et al.\ 2012, arXiv:1207.6212 

\bibitem[Cs{\'a}k et al.(2014)]{csak14} Cs{\'a}k, B., Kov{\'a}cs, J., Szab{\'o}, G.~M., et al.\ 2014, Contributions of the Astronomical Observatory Skalnate Pleso, 43, 183 

\bibitem[Dong et al.(2013)]{Dong2013} Dong, S., Katz, B., \& Socrates, A.\ 2013, \apjl, 763, L2 

\bibitem[Fabrycky \& Tremaine(2007)]{FabryckyTremaine2007} Fabrycky, D., \& Tremaine, S.\ 2007, \apj, 669, 1298 

\bibitem[Faigler \& Mazeh(2011)]{Faigler2011} Faigler, S., \& Mazeh, T.\ 2011, \mnras, 415, 3921 

\bibitem[Faigler et al.(2012)]{Faigler2012} Faigler, S., Mazeh, T., Quinn, S.~N., Latham, D.~W., \& Tal-Or, L.\ 2012, \apj, 746, 185 

\bibitem[Faigler et al.(2013)]{Faigler2013} Faigler, S., Tal-Or, L., Mazeh, T., Latham, D.~W., \& Buchhave, L.~A.\ 2013, \apj, 771, 26 


\bibitem[Faigler et al.(2015)]{Faigler2015b} Faigler, S., Kull, I., Mazeh, T., et al.\ 2015, \apj, 815, 26 

\bibitem[Fischer et al.(2016)]{Fischer2016} Fischer, D.~A., Anglada-Escude, G., Arriagada, P., et al.\ 2016, \pasp, 128, 066001 

\bibitem[Gray(1992)]{gray1992} Gray, D.~F.\ 1992,The Observation and Analysis of Stellar Photospheres (pp.368-396), Camb.~Astrophys.~Ser., Vol.~20,, 20,  

\bibitem[Hallakoun et al.(2016)]{Hallakoun2016} Hallakoun, N., Maoz, D., Kilic, M., et al.\ 2016, \mnras, 458, 845 

\bibitem[Horne(1986)]{Horne1986} Horne, K.\ 1986, \pasp, 98, 609 

\bibitem[Husser et al.(2013)]{husser2013} Husser, T.-O., Wende-von Berg, S., Dreizler, S., et al.\ 2013, \aap, 553, A6 



\bibitem[Imbert(1972)]{imbert72} Imbert, M.\ 1972, \aap, 18, 267 

\bibitem[Kiefer et al.(2016)]{Kiefer2016} Kiefer, F., Halbwachs, J.-L., Arenou, F., et al.\ 2016, \mnras, 458, 3272 

\bibitem[Kirk et al.(2016)]{Kirk2016} Kirk, B., Conroy, K., Pr{\v s}a, A., et al.\ 2016, \aj, 151, 68 

\bibitem[Kiseleva et al.(1998)]{Kiseleva1998} Kiseleva, L.~G., Eggleton, P.~P., \& Mikkola, S.\ 1998, \mnras, 300, 292 

\bibitem[Latham et al.(1989)]{Latham89} Latham, D.~W., Stefanik, R.~P., Mazeh, T., Mayor, M., \& Burki, G.\ 1989, \nat, 339, 38 


\bibitem[Loeb \& Gaudi(2003)]{LoebGaudi2003} Loeb, A., \& Gaudi, B.~S.\ 2003, \apjl, 588, L117 

\bibitem[Ma et al.(2016)]{Ma2016} Ma, B., Ge, J., Wolszczan, A., et al.\ 2016, \aj, 152, 112 

\bibitem[Mayor \& Queloz(1995)]{MayorQueloz95} Mayor, M., \& Queloz, D.\ 1995, \nat, 378, 355 

\bibitem[Mayor et al.(2014)]{Mayor2014} Mayor, M., Lovis, C., \& Santos, N.~C.\ 2014, \nat, 513, 328 


\bibitem[Mayor \& Mazeh(1987)]{MM87} Mayor, M., \& Mazeh, T.\ 1987, \aap, 171, 157 

\bibitem[Mazeh \& Shaham(1976)]{MazehShaham1976} Mazeh, T., \& Shaham, J.\ 1976, \apjl, 205, L147 

\bibitem[Mazeh \& Shaham(1979)]{MazehShaham1979} Mazeh, T., \& Shaham, J.\ 1979, \aap, 77, 145 

\bibitem[Pickles(1998)]{Pickles1998} Pickles, A.~J.\ 1998, \pasp, 110, 863 


\bibitem[Pribulla et al.(2015)]{pribulla15} Pribulla, T., Garai, Z., Hamb{\'a}lek, L., et al.\ 2015, Astronomische Nachrichten, 336, 682 



\bibitem[Tal-Or et al.(2015)]{Tal-Or2015} Tal-Or, L., Faigler, S., \& Mazeh, T.\ 2015, \aap, 580, A21 


\bibitem[Tonry \& Davis(1979)]{tonry79} Tonry, J., \& Davis, M.\ 1979, \aj, 84, 1511 

\bibitem[Zucker et al.(2007)]{Zucker2007} Zucker, S., Mazeh, T., \& Alexander, T.\ 2007, \apj, 670, 1326 




\end{thebibliography}
%



\end{document}